# Motion Parallax is Asymptotic to Binocular Disparity

by
Keith Stroyan
Mathematics Department
University of Iowa
Iowa City, Iowa

## Abstract

Researchers especially beginning with (Rogers & Graham, 1982) have noticed important psychophysical and experimental similarities between the neurologically different motion parallax and stereopsis cues. Their quantitative analysis relied primarily on the "disparity equivalence" approximation. In this article we show that retinal motion from lateral translation satisfies a strong ("asymptotic") approximation to binocular disparity. This precise mathematical similarity is also practical in the sense that it applies at normal viewing distances. The approximation is an extension to peripheral vision of (Cormac & Fox's 1985) well-known non-trig central vision approximation for binocular disparity. We hope our simple algebraic formula will be useful in analyzing experiments outside central vision where less precise approximations have led to a number of quantitative errors in the vision literature.

## Introduction

A translating observer viewing a rigid environment experiences motion parallax, the relative movement on the retina of objects in the scene. Previously we (Nawrot & Stroyan, 2009), (Stroyan & Nawrot, 2009) derived mathematical formulas for relative depth in terms of the ratio of the retinal motion rate over the smooth pursuit eye tracking rate called the motion/pursuit law. The need for inclusion of the extra-retinal information from the pursuit eye movement system for depth perception from motion parallax is now established by both human psychophysics (Nawrot, 2003), (Nawrot & Joyce, 2006), (Nawrot & Stroyan, 2009) and by primate neural recordings (Nadler, Angelaki & DeAngelis, 2008), (Nadler, Nawrot, Angelaki & DeAngelis, 2009). However, it is still interesting to compare the retinal motion cue to the binocular disparity cue, since there is a large psychophysically important literature based on the similarity of disparity and motion parallax, for example, (Gibson 1950, 1955, 1959), (Hershberger, & Starzec, 1974), (Nakayama & Loomis 1974), (Nagata 1975, 1981, 1991), (Regan & Beverly 1979, Regan 1986), (Rogers & Graham, 1979, 1982, 1984), (Richards 1985), (Ono, Rivest, & Ono 1986), (Bradshaw & Rogers 1996), (Bradshaw, Parton, & Eagle 1998), (Bradshaw, Parton, & Glennerster 2000), (Ujike & Ono 2001), (Hillis & Banks 2001), (Hanes, Keller, McCollum, 2008). Quantitative works of (Nagata 1991) and (Richards 1985) related to our result are discussed below.

The "similarity" of disparity and motion parallax is good intuition with many interesting experiments, but here we prove a corresponding mathematical similarity as a "strong" approximation "in the limit" of long fixate distances. This "asymptotic" approximation relates the geometric cues very precisely and extends the formulas into the fixation plane beyond central vision. We hope the mathematical similarity will be helpful in the analysis of future experimental investigations. Our result has no direct empirical meaning, but could be helpful in quantitative analysis of experiments as described in the discussion below.

Crossing position retinal motion is invariant on circles similar to the well-known Vieth-Müller circle of binocular disparity. These circles let us extend the asymptotic result, given only for central vision in (Nawrot & Stroyan, 2009), to the horizontal plane. The invariant circles are also a way to "compute" by relating points in peripheral vision to equivalent points in central vision. For example, the distractor $D$ in Figures 4 and 5 have the same retinal motion as the symmetric point on the dotted circle and the same disparity as the symmetric point on the dashed circle through $D$.

There is also a large literature on the related topic of "optic flow," for example, (Koenderink & van Doorn, 1975, 1976, 1987), (Longuet-Higgins & Prazdny, 1980), (Fermüller & Aloimonos, 1993, 1997), but here our result applies only to lateral motion. Fermüller & Aloimonos, (1997) study the retinal projection of "iso motion" and the horopter under optic flow, but in our case, their work only says the horopter projects onto the translation axis. Our invariant circles lie in space, their iso motion curves lie on the retina.

Often optic flow emphasizes "straight ahead" motion (such as landing an airplane) where "parallax" becomes "expansion." This is an important case, but different than we study here. (Much of Longuet-Higgens & Prazdny, for example, requires at least a component of forward motion because they divide by "$W$", which is zero for our case of lateral motion. "Dividing by zero" seems to correspond to the impossibility of "looking down your interocular axis" but there is really no impossibility except the division step in the model.) Fermüller & Aloimonos, (1993) add fixation to optic flow to solve several navigational tasks, but the optic flow literature has not to date used an appropriate vector version of the motion/pursuit ratio to give a depth formula, although there are many computational solutions to "structure from motion" based on inputs that may or may not be used by humans (for example, see Forsyth & J. Ponce, Chapters 12 & 13.) Note that (Longuet-Higgins & Prazdny, 1980), p. 391, write, "...solving such equations is not, of course, to imply that the visual system performs such calculations exactly as a mathematician would..." Stroyan & Nawrot (2009) give a partial solution to "structure from motion" based only on the psychophysically and neurologically known cues of "motion" and "pursuit." We illustrate below that the motion/pursuit depth formula follows from our asymptotic approximation (in a simple case), so the mathematical "similarity" extends to the depth formula.

## Mathematical Formulation

A common setting generating motion parallax is similar to looking out of the side window of a car as diagrammed in Figure 1. We use a rigid 2D coordinate system for the horizontal plane with one axis represented left to right on the page and the other up and down as shown in Figure 1. Our observer translates along the left-to-right "translation axis" with the right eye node passing the central intersection point of the two axes at $t = 0$. When the eye is in the central position, the nasal-occipito axis coincides with our rigid up-and-down "fixate axis." The fixate point is the 2D vector with coordinates $F = \{0, f\}$.

The angle $\alpha$ from the nasal-occipito axis to the line from eye to fixate measures the eye tracking the fixate and its rate of change with time, $d\alpha/dt$, corresponds to pursuit eye movement. (We assume the naso-occipito axis remains up-and-down on the page. At the central crossing point we have $\alpha = 0$, but observer translation forces the eye to rotate to a non-zero $\alpha$ to maintain fixation on $F$ with the nasal-occipito axis still aligned in same direction.)

The angle between the line to the fixate and the line to the distractor $\theta$ measures the separation of the images of fixate and distractor on the retina and its variation with time, $d\theta/dt$, corresponds to retinal image motion.

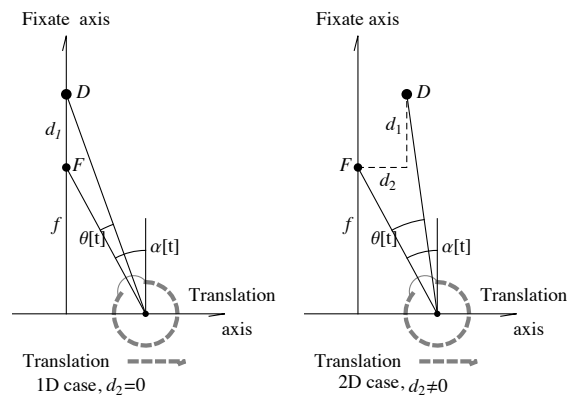

Figure 1: The tracking angle, $\alpha[t]$, separation angle, $\theta[t]$, and coordinates

Distractor points in the horizontal plane are also represented as 2D vectors in these coordinates, but we specifically represent the distractor by $D = \{d_2, f + d_1\}$, so that the vector difference $D - F = \{d_2, d_1\}$ gives the location of the distractor relative to the fixate as shown on the right portion of Figure 1. When $d_2 = 0$, the distractor lies on the fixate axis as shown on the left of Figure 1. This articles extends our prior one-dimensional analysis (described below) to include the two dimensions of the plane determined by the eyes and fixate (usually the horizontal plane). (The 1D case is only useful for distractors near central vision as we show below.)

We use the term binocular disparity to precisely mean the difference in the angles between the lines to the eyes, $B.D. = \delta = \kappa_F - \kappa_D$, where the convergence angles $\kappa_F$ and $\kappa_D$ are shown in Figure 2. Simple geometry shows that binocular disparity is also the signed difference of the separation angles $\theta$ measured counterclockwise from distraction to fixate for the two eyes, $\delta = \kappa_F - \kappa_D = \theta_r - \theta_l$. With these sign conventions, the $\theta$ formula works in all the "crossed" and "uncrossed" cases. (See (Cormack & Fox, 1985) for the complications of "crossed" and "uncrossed" caused by using a different angle convention.)

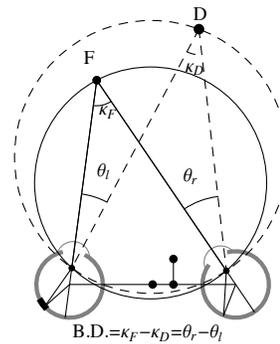

Figure 2: The left and right separation angles $\theta[t]$ and binocular disparity.

We are using the term "binocular disparity" to emphasize that our geometric measurements are made external to the eye. The disparity of the images on the actual retina depends on factors internal to the eye, especially the location of the nodal points, even if the cross section of the retina is a perfect circle as in the model of Figure 2. The internal eye parameters are separate considerations from the external geometry and we only consider the external geometry in our derivations. (The interactive program (Stroyan, 2008.7) allows the user to vary parameters and fixate and distraction locations and dynamically compute disparity.)

The dashed circle of Figure 2 drawn through the distractor and node points plays a role in our analysis below. With a single fixate, all points on the dashed circle have the same binocular disparity. In the case where the dark circle and dashed one coincide, disparity is zero and the circle is called the "theoretical horopter" or "Veith-Müller circle." (The empirical horopter has a different shape see (Hillis & Banks, 2001). Again, we only consider external geometry.) The interactive program (Stroyan, 2008.8) allows the user to vary parameters and show the invariant circles. (The program notes include a proof that binocular disparity is constant on these circles. This invariance is true for values other than disparity zero. The dark circle changes when the fixate is moved if the nodes are not at the center of the eye, because fixation changes the distance between the nodes slightly in that case.)

In this article we show that center point retinal motion ($t = 0$) is "asymptotic" to binocular disparity as the fixate distance tends to infinity. Mathematicians use "asymptotic" approximation "a~b" to mean $a/b \to 1$ or equivalently, $a/b \approx 1$ in some limit. For example, one might say the tangent of a small angle (in radians) is near the angle $\text{Tan}[\beta] \approx \beta$, but a more precise statement is that they are asymptotic, $\text{Tan}[\beta]\sim\beta$. Asymptotic means $\lim_{\beta \to 0} \text{Tan}[\beta]/\beta = 1$, so $\beta$ approximates $\text{Tan}[\beta]$ as a percentage of that small quantity. In this case both $\text{Tan}[\beta] \approx 0$ and $\beta \approx 0$, so to say they are approximately equal is only to say they are both small. A small angle $\beta$ and its square, $\beta^2$, are both approximately zero, but not asymptotic. We prove a precise mathematical theorem below, but the intuitive idea is that quantities are a strong *percentage* approximation when $f$ is large. In precise mathematics, the ratio is infinitely near 1 when $f$ is unlimited, or the ratio tends to 1 in the limit as $f$ tends to infinity. In practice the approximation is useful an "normal" viewing distances as described below.

**Theorem: Central Retinal Motion/Speed ~ Binocular Disparity/Interocular Distance**

*If the fixate distance $f$ is large and the distractor is within 45 ° nasal angle, at least a fixed percent of $f$ from the eye, and not near the Veith-Müller circle, then an observer moving perpendicular to the line to the fixate when $t = 0$ has retinal motion asymptotic to the multiple of binocular disparity in the formula:*

$$\frac{d\theta}{dt}[0] \bigg/ \left( \frac{\text{speed}}{\text{interocular distance}} \times \text{binocular disparity} \right) \approx 1 \tag{1}$$

*We also mention the simple asymptotic relation between pursuit and fixate convergence:*

$$\frac{d\alpha}{dt}[0] \bigg/ \left( \frac{\text{speed}}{\text{interocular distance}} \times \text{fixate convergence} \right) \approx 1 \tag{2}$$

The basic asymptotic comparison, $B.D. \sim d\theta/dt$, is for speed equal to the interocular distance per unit time ($\approx 6.5$ *cm/sec*). Increased speed makes the retinal motion $d\theta/dt$ larger so retinal motion can be registered by the visual system when binocular disparity is sub-threshold. (Nagata, 1991) writes, "... in the case of moving vision on some riding machine with a speed higher than the motion of the body, the sensitivities of motion parallax at large distances are maintained at that same level as that for short distances and are higher than that for binocular parallax. ..." (p.534). For example, the binocular disparity of an object 1 *m* beyond a fixate at 100 *m* is $B.D. \approx 0.02$ *min of arc (1.2 arc-sec)*, while an observer moving at 100 *km/hr* perpendicular to the line through these objects has $d\theta/dt \approx 9.5$ *min of arc/sec*. An object at these distances is above the motion parallax threshold of Nagata's Figure 3 for short distances. (Also see the less quantitative (Cutting and Vishton 1995)). (Coutant & Westheimer 1993) report that, "... At least 80% could detect depth differences at 30 sec arc disparity. ...". (Blakemore 1970) reports larger thresholds, but notes discrimination differences outside central vision. Around 1900, Bourdon, Heine, and Pulfrich (see Helmholtz 1962, p.375) reported binocular disparity discrimination in the 5 to 13 *arc-sec* range. (van der Willigen, Frost, & Wagner 1998) report 2 *min of arc* for owls and monkeys. Under very good conditions, stereo thresholds below 0.1 *min of arc* (6 *arc-sec*) are reported such as in (Schor, Bridgeman, & Tyler 1983, Figure 4), also see (Tyler 2004). In any case, 0.02 *min of arc* disparity is very small and likely can not be perceived in "normal" conditions. The binocular disparity of an object 10 *m* beyond a fixate at 100 *m* is $B.D. \approx 0.20$ *min of arc* which is near "normal" threshold, while an observer moving at 100 *km/hr* perpendicular to the line through these objects has $d\theta/dt \approx 87$ *min of arc/sec*, above Nagata's extrapolation even for "motion of the body." These comments are only intended to put the numerical speed scaling example in context, we are not making an empirical claim - just the observation that motion parallax scales with observer speed and binocular disparity does not. (Interesting work such as (Regan & Beverly 1979), (Bradshaw, Parton, & Glennerster 2000), & (Richards 1985), mentioned below, study combined stereo and motion.)

## The Motion/Pursuit Formula as a Limit of Static Depth

We begin by showing that the above asymptotic result for retinal motion gives the motion/pursuit law as the limit of the static depth formula. For simplicity, we do this for the 1D case (central vision). The exact formula for depth when the fixate and distraction are on the fixate axis (including sign) from $B.D. = \delta$ (in radians) for a given fixate distance $f$ is given by Equation (3) with $i =$interocular distance:

$$d = f \frac{4 + \left(\frac{i}{f}\right)^2}{2\left(\frac{i}{f} - 2\operatorname{Tan}[\delta/2]\right)} \operatorname{Tan}[\delta/2] \tag{3}$$

Binocular disparity alone does not determine depth since it requires the absolute fixate distance $f$. (See interactive programs (Stroyan, 2008.9 & 2008.12) that allow you to vary depth along a curve of constant B.D. or motion.) The value of $f$ could come from a vergence cue since the angle $\kappa = \kappa_F$ is related to fixate distance by $f = i/(2\operatorname{Tan}[\kappa/2])$, so (after simplification) depth from static cues is given by Equation (4).

$$d = \frac{i}{2} \frac{\operatorname{Sin}\left[\frac{\delta}{2}\right]}{\operatorname{Sin}\left[\frac{\kappa}{2}\right] \operatorname{Sin}\left[\frac{\kappa-\delta}{2}\right]} \tag{4}$$

If $i/f \approx 0$, then $\kappa \approx 0$, $\kappa/2 \sim \operatorname{Tan}[\kappa/2] \sim \operatorname{Sin}[\kappa/2]$. If the distractor is at least a real percentage of $f$, $d_1 > -k f$, for some real $k, 0 < k < 1$, then $\delta \approx 0$ and $\operatorname{Sin}[\delta/2] \sim \delta/2$, so Equation (5) gives the relative depth.

$$d = \frac{i}{2} \frac{\operatorname{Sin}\left[\frac{\delta}{2}\right]}{\operatorname{Sin}\left[\frac{\kappa}{2}\right] \operatorname{Sin}\left[\frac{\kappa-\delta}{2}\right]} \sim f \frac{\delta}{\kappa - \delta}, \text{ and the relative depth is,} \quad \frac{d}{f} \sim \frac{\delta}{\kappa - \delta} \tag{5}$$

The pursuit derivative when observer speed is $i$, the interocular distance, satisfies Equation (6).

$$\frac{d\alpha}{dt} = \frac{i}{f} \text{ (when } n = 0\text{)} \; \& \text{ since } \operatorname{Tan}\left[\frac{\kappa}{2}\right] = \frac{(i/2)}{f}, \; \frac{d\alpha}{dt}[0] = \dot{\alpha} \sim \kappa \text{ when } \frac{i}{f} \approx 0. \tag{6}$$

Using the asymptotic result $\frac{d\theta}{dt}[0] = \dot{\theta} \sim \delta$ the static relative depth approximation of Equation (6) becomes Equation (7), the 1D motion/pursuit law.

$$\frac{d}{f} \sim \frac{\delta}{\kappa - \delta} \sim \frac{\dot{\theta}}{\dot{\alpha} - \dot{\theta}} = \frac{d}{f} \tag{7}$$

Of course, (Nawrot & Stroyan, 2009) already showed that the motion/pursuit formula gives the relative depth in the 1D case. This derivation of the central vision Motion/Pursuit Law is to show that the formula could be obtained from the asymptotic relations of the theorem above. Disparity and vergence play a roles similar to motion and pursuit geometrically (if not psychophysically or neurologically. We do not make any empirical claim - just note the mathematical similarity of the pairs of terms in the static and dynamic relative depth formulas. It is curious that no one noticed this before. Perhaps this is because of the observations of (Wundt, 1861 - see: Helmholtz 1962), (Helmholtz 1962, p.315) and others that, "The uncertainty in estimating the absolute amount of convergence ... is manifested in many instances." Also, psychophysically, (Nawrot & Stroyan, 2009) showed that only the ratio $\dot{\theta}/\dot{\alpha}$ affected percepion, so perhaps formula (7) is better written, $(\dot{\theta}/\dot{\alpha})/(1 - (\dot{\theta}/\dot{\alpha}))$, and we do not propose a stereo counterpart to this ratio.)

## Formulas for Retinal Motion

To simplify the exposition we will assume that the node is at the center of the eye and static fixation is symmetric. (We have derived formulas with a variable node position and they are both considerably more complicated and approximately equal to the central node formulas. A non-central node makes the node speed slightly different from the observer's because of fixation tracking.) In (Stroyan & Nawrot, 2009) we show that for an observer translating at speed $s$ along the translation axis of Figure 1, the retinal motion is given by

$$\frac{d\theta}{dt} = \frac{s}{f^2 + s^2 t^2}\left(f - \frac{(d_1 + f)\left(f^2 + s^2 t^2\right)}{(d_1 + f)^2 + (d_2 - s t)^2}\right) \qquad (8)$$

At the central point where when the eye crosses the fixate axis when $t = 0$, this simplifies to:

$$\frac{d\theta}{dt}[0] = \frac{\left(d_1^2 + d_2^2 + d_1 f\right) s}{\left((d_1 + f)^2 + d_2^2\right) f} \qquad (9)$$

This time zero retinal motion is constant on circles with diameter on the line from the right eye node $\{0, 0\}$ with $d_1 = -f$ to the symmetric point $\left\{0, \frac{sf}{s-cf}\right\} = \{0, f + d'\}$ with $d' = \frac{sf}{s-cf} - f = \frac{cf^2}{s-cf}$. Formally, this is Equation (10) as shown on the left graph of Figure 3.

$$\frac{d\theta}{dt}[0] = c \iff \left((d_1 + f) - \frac{sf}{2(s - cf)}\right)^2 + d_2^2 = \left(\frac{sf}{2(s - cf)}\right)^2 \qquad (10)$$

When $\frac{d\theta}{dt}[0] = c$ for the $y$-axis fixate $\{0, f\}$ and distraction $\{0, f + d\}$, the diameter is from $\{0, 0\}$ to $\{0, f + d\}$ with radius $\frac{f+d}{2}$ and center $\left\{0, \frac{f+d}{2}\right\}$.

The one dimensional retinal motion derivative, $\frac{d\theta}{dt}[0]$ when the distraction is in line with the eye, $d_2 = 0$, has the simple form of Equation (11).

$$\frac{d\theta}{dt}[0] = \frac{s\, d_1}{(d_1 + f)\, f} \qquad (11)$$

## Proof of the 2D Asymptotic Result

In (Nawrot & Stroyan, 2009) we proved that $B.D. \sim \frac{d\theta}{dt}[0]$ when $\frac{i}{f} \approx 0$, $d_1 + f \neq 0$, $d_1 \neq 0$, and both **F** and **D** lie on the fixate axis, **F** at $\{0, f\}$ and **D** at $\{0, f + d_1\}$. That proof is only for the 1D case of distractors along the fixate axis, $d_2 = 0$. Now we establish the theorem above showing that $\delta/\dot{\theta} \approx 1$ in two dimensions when speed equals the interocular distance, $s = i$, (using $\dot{\theta} = d\theta/dt[0]$ for simplicity.)

Figure 3 shows graphs of both $\dot{\theta}$ and binocular disparity in 2D. If observer speed equals the interocular distance per unit time, the vertical scales of both graphs are the same. The two graphs are similar for a 2 dimensional distractor in that they are constant on slightly different circles shown on the graphs.

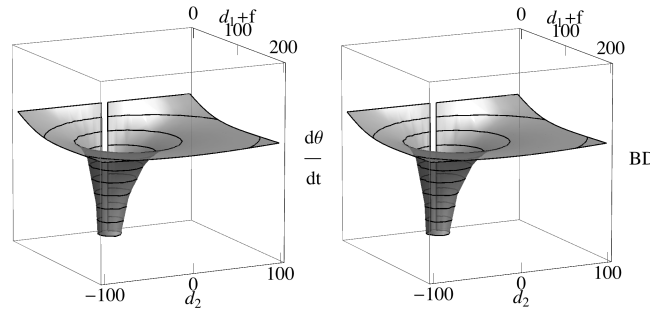

Figure 3: Two dimensional central retinal motion (left) and binocular disparity (right), $f = 100$, $s = i = 6.5$

We use the different invariant circles for both disparity and motion to move to equivalent points on the fixate axis with the same respective values. The values of the respective quantities are the same at these fixate axis points as at the original **D** and we can use our one dimensional result on the fixate axis to show that the two quantities are asymptotic at the original **D**.

The invariant circle for binocular disparity passes through the distractor and the left and right eye nodes, while the invariant circle for $t = 0$ retinal motion passes through the node of the eye as it crosses the fixate axis (and has a diameter on the $d_1$-axis). The points where the circles cross the fixate axis are shown in Figure 4. It is clear that the invariant circles tend together as the fixate distance gets long because the eye nodes converge to the central crossing point as a fraction of fixate distance. We simply calculate the exact locations of the equivalent points as follows.

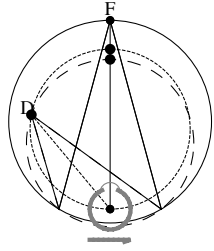

Figure 4: Circles of constant two dimensional retinal motion and binocular disparity

For retinal motion, the equivalent point on the fixate axis is at the point given in Equation (12).

$$\{0, p\} = \left\{0, (d_1 + f)\left[1 + \left(\frac{d_2}{d_1 + f}\right)^2\right]\right\} \tag{12}$$

For binocular disparity, the equivalent point on the fixate axis is given in Equation (13).

$$p' = (d_1 + f) \frac{1 + \left(\frac{d_2}{d_1+f}\right)^2 - \left(\frac{i}{2(d_1+f)}\right)^2 + \sqrt{\left(\frac{i}{(d_1+f)}\right)^2 + \left(1 + \left(\frac{d_2}{d_1+f}\right)^2 - \left(\frac{i}{2(d_1+f)}\right)^2\right)^2}}{2} \tag{13}$$

Assuming that $\frac{1}{d_1+f} \approx 0$ and $\frac{d_1+f}{d_2} \not\approx 0$ (for example if the distractor is far away and at a nasal angle less than 45°), the points $p$ and $p'$ are infinitely close by the computation in Equation (14).

$$p - p' = (d_1 + f)\left[1 + \left(\frac{d_2}{d_1 + f}\right)^2\right] - (d_1 + f) \frac{1 + \left(\frac{d_2}{d_1+f}\right)^2 - \left(\frac{i}{2(d_1+f)}\right)^2 + \sqrt{\left(\frac{i}{(d_1+f)}\right)^2 + \left(1 + \left(\frac{d_2}{d_1+f}\right)^2 - \left(\frac{i}{2(d_1+f)}\right)^2\right)^2}}{2} =$$

$$= \frac{(d_1 + f)}{2}\left(1 + \frac{d_2^2}{(d_1+f)^2} + \frac{i^2}{4(d_1+f)^2} - \sqrt{\frac{i^2}{(d_1+f)^2} + \left(1 + \frac{d_2^2}{(d_1+f)^2} - \frac{i^2}{4(d_1+f)^2}\right)^2}\right) =$$

$$= \frac{(d_1 + f)}{2} \frac{\left(1 + \frac{d_2^2}{(d_1+f)^2} + \frac{i^2}{4(d_1+f)^2}\right)^2 - \frac{i^2}{(d_1+f)^2} - \left(1 + \frac{d_2^2}{(d_1+f)^2} - \frac{i^2}{4(d_1+f)^2}\right)^2}{\left(1 + \frac{d_2^2}{(d_1+f)^2} + \frac{i^2}{4(d_1+f)^2} + \sqrt{\frac{i^2}{(d_1+f)^2} + \left(1 + \frac{d_2^2}{(d_1+f)^2} - \frac{i^2}{4(d_1+f)^2}\right)^2}\right)} =$$

$$= \frac{(d_1 + f)}{2} \frac{\frac{d_2^2 \, i^2}{(d_1+f)^4}}{\left(1 + \frac{d_2^2}{(d_1+f)^2} + \frac{i^2}{4(d_1+f)^2} + \sqrt{\frac{i^2}{(d_1+f)^2} + \left(1 + \frac{d_2^2}{(d_1+f)^2} - \frac{i^2}{4(d_1+f)^2}\right)^2}\right)} =$$

$$= \frac{1}{2(d_1+f)}\left(\frac{d_2\,i}{d_1+f}\right)^2 \frac{1}{\left(1+\left(\frac{d_2}{d_1+f}\right)^2+\left(\frac{i}{2(d_1+f)}\right)^2+\sqrt{\left(\frac{i}{d_1+f}\right)^2+\left(1+\left(\frac{d_2}{d_1+f}\right)^2-\left(\frac{i}{2(d_1+f)}\right)^2\right)^2}\right)} \approx 0$$

$$p - p' = \frac{1}{(d_1+f)} q \approx 0, \text{ for limited } q. \tag{14}$$

We know from the 1D case above that $B.D.[p'] \big/ \frac{d\theta}{dt}[p'] = 1 + \varepsilon$, with $\varepsilon \approx 0$. The derivative in the direction of the fixate axis is

$\frac{\partial d\theta/dt}{\partial d} = \frac{i}{(d+f)^2} = \frac{f}{d(d+f)} \frac{d\,i}{(d+f)\,f} = \frac{f}{d(d+f)} \frac{d\theta}{dt}[p]$, so we have Equation (15).

$$\frac{d\theta}{dt}[p'] = \frac{d\theta}{dt}[p]\left(1 + \frac{1}{(d_1+f)} \frac{f}{d_1(d_1+f)} q'\right), \text{ for limited } q' \tag{15}$$

As long as $d_1 \not\approx 0$ and $d_1 > -k\,f$, for some real $k$, $0 < k < 1$, then $f/(d_1+f) < 1/(1-k)$, and this gives Equation (16).

$$B.D.[\{d_1, d_2\}] \Big/ \frac{d\theta}{dt}[p'] = B.D.[p'] \Big/ \frac{d\theta}{dt}[p'] \approx B.D.[\{d_1, d_2\}] \Big/ \frac{d\theta}{dt}[p] = B.D.[\{d_1, d_2\}] \Big/ \frac{d\theta}{dt}[\{d_1, d_2\}] \approx 1 \tag{16}$$

This gives us the main result, formula (1) above. This theorem could be stated as the limit of the ratio tending to 1 as $f \to \infty$ for a compact set of distractors, but we leave that formulation to the reader. (See (Stroyan, 1998) or other references on Robinson's theory of infinitesimals. Intuitively, $\approx$ just means "very close.")

Perhaps we should point out that there is nothing "wrong" near the Veith-Müller circle. Both quantities are near zero and trying to compare them as a ratio does not work in that case. Increasing the node distance from the center of the eye makes a slightly larger separation between the Veith-Müller circle and the zero retinal motion circle, but the fundamental approximation still holds. Also, the central position of motion parallax is not essential. Motion at the right eye would work, but give a slightly less accurate approximation at a given $f$.

**Comparison with Retinal Motion in Older Work**

This asymptotic approximation is a 2D extension of the well-known 1-D formulas of (Cormack & Fox, 1985) or (Davis & Hodges, 1995). They derived the approximation by trig approximations without noting that the 1-D algebraic formula is the center line retinal motion rate (when speed equals the interocular distance per second). In particular, they gave the "nontrig law assuming symmetric fixation" (formula (11) of (Cormack & Fox, 1985) re-written in our notation for distances (the same as formula (4) of (Davis & Hodges, 1995)) as Equation (17).

$$B.D. \sim \frac{i}{f} \frac{d}{(f+d)} \tag{17}$$

The right hand side of the approximation is the expression for the retinal motion rate in formula (11) above when $s = i$, in other words, (Cormack & Fox, 1985) formula (11) is the 1-D approximation $B.D. \sim \frac{d\theta}{dt}$ although neither (Cormack & Fox, 1985) nor (Davis & Hodges, 1995) notice that the algebraic expression is the retinal motion rate.

A two dimensional extension of (Cormack & Fox, 1985) formula (11) is given by the expression for retinal motion at time zero, Equation (18).

$$B.D.[d_1, d_2] \sim \frac{d\theta}{dt}[0] = \frac{i}{f} \frac{(f+d_1)\,d_1 + d_2^2}{(f+d_1)^2 + d_2^2} \tag{18}$$

The formula may look complicated, but one can not make it simpler for peripheral vision. For example, if $f = 57\ cm$, $d_1 = 2\ cm$, $d_2 = -6\ cm$, then B.D.= 17.10 *arc minutes* and formula (18) = $d\theta/dt[0]$ = 17.17, a 0.4% error, but formula (17) = 13.29, a 22.3% error. See Figure 5.a. One can accurately approximate binocular disparity by using the central calculations and moving around the invariant circle. In the example of Figure 5.a the equivalent points are at 59.608 *cm* and 59.610 *cm*. The interactive program of (Stroyan, 2008.14) lets the user vary the position of fixate and distraction and dynamically computes these quantities. The approximation is accurate at "normal" viewing distances, not only in the limit as $f \to \infty$.

## Discussion

While the derivation of our asymptotic approximation formula is quite technical, the idea is simple. Both binocular disparity and crossing time retinal motion are invariant on similar circles. We compare 2D binocular disparity and 2D retinal motion by computing the 1D case with the fixate and distractor on the fixate axis at equivalent symmetric points of the appropriate invariant circles shown in Figure 4. The calculations show that these equivalent points are close when the fixate distance is long, so the two quantities are "asymptotically" close. "Obvious," but the specifics give a strong approximation and geometric "similarity."

Our result is important as an approximation only when $d_2$ is not negligible, corresponding to points outside central vision. We have noticed a number of analytical errors in the vision literature in cases where $d_2$ should not be neglected. We do not want to over-emphasize the errors, papers such as Hanes, Keller & McCollum (2008) also point out the basic difficulty, but these errors do occur (such as in the very interesting work of Richards (1985) which has an incorrect Figure 1.) Without making a long excursion into specifics, the difficulties arise in situations illustrated in Figure 5.b. The "square law," $i\,d_1/f^2$, in this illustrative example gives disparity (or motion) zero for the laterally displaced distractor shown, but in the example we have $f = 57cm$, $d_1 = 0$, and $d_2 = -15$, so the disparity is actually 25.24 *arc-minutes* and formula (18) gives 25.39. This is far from zero. The equivalent symmetric points are at 60.94 and 60.95 *cm*. This same basic difficulty arises in Nagata's (1991) analysis of the threshold of perception of motion parallax. In his experiments, the distractor is off-set from the fixate, so his claimed threshold may be too low. We do not have enough of the details of his experiments to propose an adjustment, nor do we herewith propose an alternative to Richards structure from motion estimates. They are interesting papers, but might each deserve a second look in quantitative analysis.

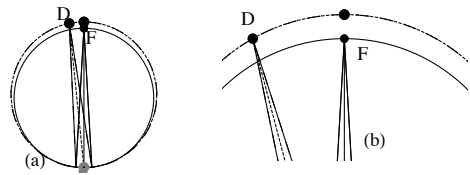

Figure 5: Non-zero retinal motion and binocular disparity outside central vision

"Disparity equivalence" is a simple intuitive approximation that amounts to calculating the difference in the view of a moving eye between the static positions of the left and right eye. It's fine, it's simple, but it is not as precise as our approximation and it does not show the geometric "similarity" of the cues that our "circle analysis" does. We consider our approximation (18) mainly a statement of geometric "similarity" and a useful approximation formula for binocular disparity outside central vision. Disparity equivalence can also be considered an average of the instantaneous retinal motion over the period it takes the observer to travel the interocular distance. One might object to an "instantaneous" perceptual cue, so some "integration time" might be called for in using $d\theta/dt$ experimentally. One can simply average formula (8) over short time intervals. This would allow an experimenter to account for the empirical integration time, but our preliminary computations for "reasonable" integration times and "normal" viewing distances indicate that formula (18) is only changed slightly by averaging.